%% file: Prasobh_ISAC_DL_v1.tex
\documentclass[journal]{IEEEtran}
\IEEEoverridecommandlockouts
\usepackage{cite}
\usepackage{amsmath,amssymb,amsfonts}
\usepackage{graphicx}
\usepackage{acronym,comment}
\usepackage[colorinlistoftodos]{todonotes}

\usepackage{textcomp}
\usepackage{xcolor}

\usepackage{bbm}
\usepackage{amsthm}
\usepackage{psfrag,caption,subcaption}

\usepackage{bbm}
\usepackage{algorithm}
\usepackage[noend]{algpseudocode} 

\algrenewcommand\algorithmiccomment[1]{// {\itshape #1}}

\usepackage[utf8]{inputenc}
\usepackage[english]{babel}
\newtheorem{theorem}{Theorem}
\input{macros}

\acrodef{ris}[RIS]{reconfigurable intelligent surface}
\acrodef{isac}[ISAC]{integrated sensing and communication}
\acrodef{dfbs}[DFBS]{dual function radar communication base station}
\acrodef{ue}[UE]{user equipment}
\acrodef{sinr}[SINR]{signal-to-interference-plus-noise ratio}
\acrodef{snr}[SNR]{signal-to-noise ratio}
\acrodef{mui}[MUI]{multi-user-interference}
\acrodef{mimo}[MIMO]{multiple-input-multiple-output}
\acrodef{miso}[MISO]{multiple-input-single-output}
\acrodef{ura}[URA]{uniform rectangular array}
\acrodef{nlos}[NLoS]{non-line-of-sight}
\acrodef{los}[LoS]{line-of-sight}
\acrodef{wrt}[w.r.t.]{with respect to}
\acrodef{rcc}[RCC]{radar-communication-coexistence}
\acrodef{rcs}[RCS]{radar cross section}
\acrodef{ula}[ULA]{uniform linear array}
\acrodef{bs}[BS]{base station}
\acrodef{qos}[QoS]{quality of service}
\acrodef{cdf}[c.d.f.]{cumulative distribution function}
\acrodef{sdp}[SDP]{semidefinite program}
\acrodef{psd}[PSD]{positive semi definite }
\acrodef{iid}[IID]{independent and identically distributed}
\acrodef{csi}[CSI]{channel state information}
\acrodef{ls}[LS]{least squares}
\acrodef{nn}[NN]{neural network}
\acrodef{dl}[DL]{deep learning}
\acrodef{mlp}[MLP]{multi-layer perceptron}
\acrodef{sota}[SoTA]{state-of-the-art}
\acrodef{lmmse}[LMMSE]{linear minimum mean squared error}
\acrodef{dris}[DP-HRIS]{dynamically programmable hybrid reconfigurable intelligent surface}

\def\BibTeX{{\rm B\kern-.05em{\sc i\kern-.025em b}\kern-.08em
		T\kern-.1667em\lower.7ex\hbox{E}\kern-.125emX}}
\begin{document}
	
	\title{
		\color{black} Learning to Precode for Integrated \\ Sensing and Communications Systems
	}
	
	\author{\IEEEauthorblockN{ R.S. Prasobh Sankar, Sidharth S. Nair, Siddhant Doshi, and  Sundeep Prabhakar Chepuri \\ }
		\IEEEauthorblockA{ {Indian Institute of Science,}
			Bengaluru, India  
		}
	}
	\maketitle
	
	\begin{abstract}
In this paper, we present an unsupervised learning neural model to design transmit precoders for integrated sensing and communication (ISAC) systems to maximize the worst-case target illumination power while ensuring a minimum signal-to-interference-plus-noise ratio (SINR) for all the users. The problem of learning transmit precoders from uplink pilots and echoes can be viewed as a parameterized function estimation problem and we propose to learn this function using a neural network model. To learn the neural network parameters, we develop a novel loss function based on the first-order optimality conditions to incorporate the SINR and power constraints. Through numerical simulations, we demonstrate that the proposed method outperforms traditional optimization-based methods in presence of channel estimation errors while incurring lesser computational complexity and generalizing well across different channel conditions that were not shown during training.

	\end{abstract}
	
	\begin{IEEEkeywords}
		Beamforming, integrated sensing and communication, neural network, precoding, unsupervised learning.
	\end{IEEEkeywords}

\section{Introduction} \label{sec:intro} 

\Ac{isac} systems such as \acp{dfbs} that carry out both communication and sensing while sharing hardware and spectral resources are expected to play a key role in the next generation of wireless systems~\cite{liu2022ISAC_6G,chepuri2022ISACwithRIS}. 
A careful design of transmit precoders is essential  to fully  utilize the available degrees of freedom and to achieve good tradeoff between the sensing and communication performance in \ac{isac} systems~\cite{liu2020joint_transmit_beamform,hua2022optimalISAC}.

{\color{black}Typically, transmit precoders are designed to maximize a radar metric~(e.g., target illumination power or the received  radar \ac{snr}) while guaranteeing a \ac{qos} metric~(e.g., \ac{sinr} or rate) for the communication users, or vice versa}. Assuming perfectly known wireless channels,  transmit precoders are  designed by solving a~\ac{sdp} optimization~\cite{liu2020joint_transmit_beamform,hua2022optimalISAC}. In practice, the wireless channels are estimated prior to the precoder design. {\color{black}In presence of channel estimation errors, the transmit precoders obtained using methods requiring perfect \ac{csi} will also be erroneous.  Further, the use of \acp{sdp} are computationally intensive.}

Learning-based methods to solve optimization problems in wireless systems have been receiving steady attention~\cite{sun2018learningtooptimize,balevi2020Massive_MIMO_channel_estm_DL,jiang2021learning_to_reflect}, wherein instead of using traditional optimization based methods, \acp{nn} are used to learn the solution to the underlying optimization problem. Learning-based methods have been found  beneficial for transmit beamforming~\cite{sun2018learningtooptimize}, channel estimation~\cite{balevi2020Massive_MIMO_channel_estm_DL}, and joint beamforming and reflection design~\cite{jiang2021learning_to_reflect} in communication systems, to name a few. Leveraging the ability of \acp{nn} to learn functions, it has been shown that  \acp{nn} can be trained in an unsupervised setting to learn  transmit precoders directly from the received pilots at the \ac{bs} while completely bypassing the explicit channel estimation stage for \ac{mimo} communication systems~\cite{jiang2021learning_to_reflect}. Nevertheless, \cite{jiang2021learning_to_reflect} is limited to  non-\ac{isac} scenarios where the underlying optimization problem comprises of simple constraints, such as the total power constraint, which can be ensured via a simple normalization layer in \acp{nn}.  Since \ac{isac} systems typically involve complex constraints, e.g., \ac{sinr} constraints, which cannot be ensured via simple scaling operations, a direct application of~\cite{jiang2021learning_to_reflect} for \ac{isac} systems is non-trivial.

In this paper, we present an \ac{nn}-based  approach to design  transmit precoders in an \ac{isac} system wherein the transmit precoders are designed by maximizing the worst-case target illumination power subject to per user \ac{sinr} constraints and transmit power constraint. We consider  the transmit precoder design problem as that of learning a function that maps the set of received communication pilots and echoes to the set of admissible precoders. We learn the mapping using an \ac{nn}. Specifically, we present an \ac{nn} architecture that takes communication pilots and radar echoes as inputs and outputs the \ac{isac} transmit precoder without explicitly estimating or knowing \ac{csi} or target locations. We train the \ac{nn} in an unsupervised setting using a loss function that maximizes the radar objective while promoting solutions that satisfy communication constraints. To train the network with \ac{sinr} and power constraints, we propose a loss function based on the first-order optimality conditions to jointly learn both the weights of the \ac{nn} and the Lagrange multipliers.   Through numerical simulations, we demonstrate that the proposed approach ensures the required minimum \ac{sinr} for all the \acp{ue} in a stochastic sense (i.e., on an average across multiple channel realizations)  while having superior sensing performance when compared to applying traditional optimization-based methods on estimated channels. The proposed \ac{nn} model also generalizes well across different \ac{ue} locations and number of \acp{ue}. The proposed method also offers significantly lower complexity when compared with \ac{sdp}-based methods~\cite{liu2020joint_transmit_beamform} and scales linearly with the number of \acp{ue}, making it suited for massive \ac{mimo} \ac{isac} systems.

\section{System model and transmit beamforming}
In this paper, we consider an \ac{isac} system serving $K$ single antenna \acp{ue} and sensing $T$ targets. 
We model the \ac{dfbs} as a \ac{ula} of $M$ antennas with half-wavelength spacing.

\subsection{Downlink transmit signal}
The \ac{dfbs} transmits a superposition of communication symbols $\vd_n\in \mbC^K$ and sensing waveforms $\vt_n \in \mbC^M$. Specifically, the communication symbols and the sensing waveforms are precoded with the communication precoder $\mC = [\vc_1,\ldots,\vc_K] \in \mbC^{M \times K}$ and the sensing precoder $\mS = [\vs_1,\ldots,\vs_M] \in \mbC^{M \times M}$, respectively. 
We assume that the transmit precoders satisfy a total power constraint of $P_{\rm d}$, i.e., $\Vert \mW \Vert_F^2 = P_{\rm d}$, where $\mW = [\mC,\mS] \in \mbC^{M \times (M+K)}$ is the overall transmit precoder.  The overall downlink transmit signal is
\begin{eqnarray}
	\vx_n =  \mC\vd_n + \mS \vt_n .
\end{eqnarray}

Let $\vh_k\rH \in \mbC^{1\times M}$ denote the \ac{miso} channel from the \ac{dfbs} to the $k$th \ac{ue}. The signal received at the $k$th \ac{ue} is given by
\begin{equation}
	y_{k,n}^{\rm dl} = \vh_k\rH \vx_n + e_{k,n},
\end{equation}
where $e_{k,n} \sim \cC\cN(0,\sigma^2)$ is the receiver noise. The corresponding~\ac{sinr} is given by
\begin{equation}
	\gamma_k\left( \mW\right) = \frac{ \vert \vh_k \rH \vc_k \vert^2 }{ \sum_{ j=1,j\neq k}^{K} \vert  \vh_k \rH \vc_j \vert^2 + \sum_{m=1}^{M} \vert \vh_k \rH \vs_m \vert^2 + \sigma^2 }.
\end{equation}

We model each target as a point scatterer present in the far field of the \ac{dfbs} with the $m$th target present at angle of $\theta_m$ \ac{wrt} the \ac{dfbs}. Let us define the channel corresponding to the $m$th target as $\vg_m\rH = \alpha_m \va\rT(\theta_m)$,
where $\alpha_m$ is the overall fading coefficient and  $\va(\theta) = [1,e^{-\jmath \pi \sin \theta},\ldots,e^{-\jmath \pi (M-1) \sin \theta}]\rT \in \mbC^M$ denotes the array response vector of the \ac{ula} at the \ac{dfbs} towards the direction $\theta$. The echo received  at the \ac{dfbs} after getting reflected from the targets is given by
\begin{equation} \label{eq:rec:echo}
	\vz_n = \sum_{m=1}^T \beta_m \vg_m^{*}\vg_m\rH  \vx_{n-n_{0,m}} + \vv_{n},
\end{equation}
where $\beta_m$ is the \ac{rcs} of the $m$th target, $n_{0,m}$ is the discrete-time round trip delay (a function of the range) corresponding to the $m$th target and $\vv_{n} \sim \cC\cN(\boldsymbol{0},\nu^2\mI)$ is the receiver noise at the \ac{dfbs}. In the remainder of the paper, we assume that the range of all targets are same and known (i.e., we set $n_{0,m}=0$ for  $m=1,\ldots,T$) for simplicity, and focus on the problem of sensing and beamforming towards the $T$ targets in the spatial domain. The illumination power of the $m$th target is defined as $Q_m   \overset{\Delta}{=} \mbE [  \vert \vg_m\rH\vx \vert^2  ]$ with the worst-case target illumination power being
\begin{equation} 
	Q\left( \mW \right) = \underset{1 \leq m \leq T}{\text{min}} \quad \vg_m\rH  \mW\mW\rH \vg_m.
\end{equation}
Next, we state the transmit precoding problem when the channels are perfectly known.

\subsection{Transmit beamformer design with channel knowledge}
The symbol decoding capability of a \ac{ue} in a multi-user setting is determined by the \ac{sinr}. Similarly, for well separated targets, the ability of a radar to successfully sense targets is directly proportional to the target illumination power~\cite{stoica2007on_probing_signal}. Hence,  to design the precoders, we maximize the worst-case target illumination power while guaranteeing a certain \ac{sinr}, say $\Gamma$, for each \ac{ue}. That is, we solve
\begin{subequations} \label{prob:form}
	\begin{align} 
		(\mathcal{P}):	\quad \underset{\mW}{\text{maximize}} & \quad Q(\mW)  \nonumber\\
	  \,\,	\text{s. to} &\quad {\rm Tr}\left(\mW\mW\rH \right) = P_{\rm d}, \label{prob:form:pow}\\
		& \quad \gamma_k(\mW) \geq \Gamma, \,\, k=1,\ldots,K,  \label{prob:form:sinr}
	\end{align}
\end{subequations}
where~\eqref{prob:form:pow} is the total transmit power constraint at the \ac{dfbs} and~\eqref{prob:form:sinr} is the per user fairness \ac{sinr} requirement. The problem $(\cP)$ can be solved as an \ac{sdp} with a complexity of about $\cO(M^{6.5}K^{6.5}\log (1/\delta))$ with an accuracy of $\delta$~\cite{sankar2022isac_ris_beamforming,liu2020joint_transmit_beamform}. 

\section{Learning-based transmit beamforming}
In practice, to solve ($\cP$), we have to estimate the wireless channels. The communication channels are estimated based on the uplink pilot symbols received from the \acp{ue}, e.g., using least-squares, whereas the target directions $\theta_m$ and coefficients $(\alpha_m,\beta_m)$ are estimated based on the echo signals received at the \ac{dfbs} using direction-finding methods and least-squares, respectively. The precoder design problem $(\cP)$ is then carried out using the estimated channels. However, applying traditional optimization-based techniques~\cite{liu2020joint_transmit_beamform,hua2022optimalISAC} on estimated channels leads to error propagation from channel estimates to the solution of $(\cP)$. To alleviate the error propagation, we present a learning-based framework to directly obtain the transmit precoders. To begin with, we present a sounding scheme to obtain the required data for channel estimation i.e., the pilots and the echoes. 

\subsection{Channel sounding for data acquisition}
First, the \acp{ue} transmit uplink pilots symbols to the \ac{dfbs}. Let $\mH \in \mbC^{K \times M}$ be the downlink channel matrix with $\vh_k\rH$ being the $k$th row. The \acp{ue} transmit orthogonal pilot sequence of length $L_{\rm p} \geq K$, $\mF \in \mbC^{K \times L_p}$ such that $\mF\mF\rH = L_p \mI_K$. The signal received at the \ac{dfbs} is
\begin{equation} \label{eq:rec:pilot}
	\mY = \sqrt{P_{\rm u}} \mH\rT \mF + \mN,
\end{equation}   
where $P_{\rm u}$ is the transmit power of each \ac{ue} and $\mN$ is the receiver noise at the \ac{dfbs} with $[\mN]_{i,j} \sim \cC\cN(0,\nu^2)$. Removing the known pilots yields $\tilde{\mY} = \mY\mS\rH$.

\begin{figure*}[t]
	\includegraphics[width=\linewidth]{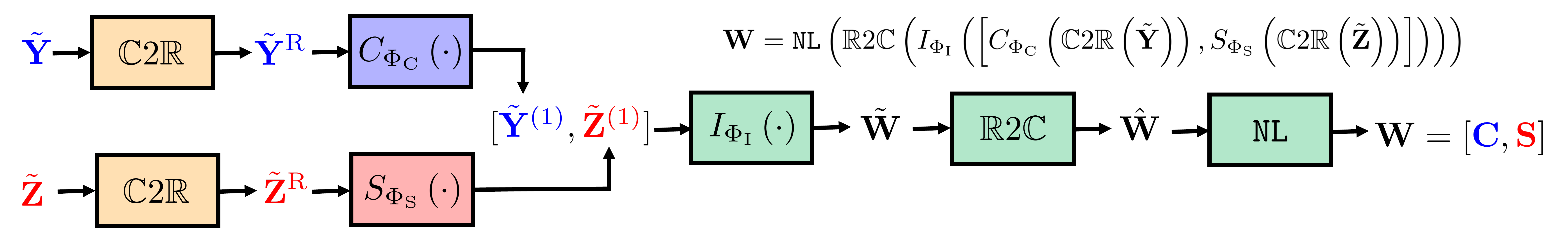}
	\caption{The proposed \ac{nn} architecture.}
	\label{fig:nn:arch}
 \vspace{-3mm}
\end{figure*}

{\color{black}
 Next, the \ac{dfbs} transmits sensing waveforms. Since the target directions are not known, the sensing waveforms are selected to uniformly illuminate all angles~\cite{stoica2007on_probing_signal}.   Let $\mE \in \mbC^{M \times L_{\rm r}}$ denote the $L_{\rm r}$-long omnidirectional waveforms transmitted from the \ac{dfbs} satisfying $\va\rH(\theta)\left( \mE\mE\rH \right)\va(\theta) = c > 0$ for all $\theta$. Without the loss of generality, we assume that the sensing waveforms $\mE$ are orthogonal with $\mE\mE\rH = (L_{\rm r}/M)\mI$ with $c = L_{\rm r}$. Using~\eqref{eq:rec:echo}, the received echo signals at the \ac{dfbs} can be written as
 \begin{equation} \label{eq:rec:echo:mat}
 	\mZ = \sqrt{P_{\rm r}}\sum_{m=1}^T \beta_m \vg_m^{*}\vg_m\rH  \mE + \mV,
 \end{equation}
 where $P_{\rm r}$ is the total transmit power of sensing waveforms and  $[\mV]_{i,j} \sim \cC\cN(0,\nu^2)$ for all $i,j$ is the additive noise at the \ac{dfbs}. 
The received signal is matched filtered to obtain $\tilde{\mZ} \overset{\Delta}{=} \mZ\mE\rH$.}
We now present the proposed learning-based formulation for the transmit precoder design. 

\vspace{-3mm}
\subsection{Learning-based formulation}
Let us denote the solution to the optimization problem $(\cP)$ by $\mW^\star$. Since the objective functions and the constraints are functions of the wireless channels, the optimal solution $\mW^\star$ is also a function of the channels, i.e., $\mW^\star = f(\mH,\vg_1,\ldots,\vg_T)$. Moreover, since the wireless channels are estimated from the pilots and echoes, the optimal solution $\mW^\star$ is a function of $\tilde{\mY}$ and $\tilde{\mZ}$, i.e., $\mW^\star = f_{\Phi^\star}( \tilde{\mY},\tilde{\mZ} )$, where the function $f_{\Phi}\left( \cdot \right) \colon \mbC^{M \times K} \times \mbC^{M \times M} \rightarrow \mbC^{M \times (M+K)}$  parameterized by $\Phi$, maps the received pilots and echoes to the solution $\mW^\star$. Hence, $(\cP)$ can be equivalently restated as the problem of finding the optimal parameter $\Phi^\star$, which is obtained by solving
\begin{subequations} \label{prob:form:learn}
	\begin{align} 
 \underset{\Phi}{\text{maximize}} & \quad Q\left( f_{\Phi}\left( \tilde{\mY},\tilde{\mZ} \right) \right) \nonumber \\
			\text{s. to} &\quad {\rm Tr}\left(f_{\Phi}\left( \tilde{\mY},\tilde{\mZ} \right)f_{\Phi}\rH\left( \tilde{\mY},\tilde{\mZ} \right) \right) = P_{\rm d} \label{prob:form:pow:learn}\\
		& \quad \gamma_k\left(f_{\Phi}\left( \tilde{\mY},\tilde{\mZ} \right)\right) \geq \Gamma, \,\, k=1,\ldots,K . \label{prob:form:sinr:learn}
	\end{align}
\end{subequations}
Next, we propose an \ac{nn}-based solution to solve~\eqref{prob:form:learn}.

\section{The proposed \ac{nn} precoder}
In this section, we propose to leverage the universal function approximation~\cite{hornik1989universalapptheo} ability of \acp{mlp} to learn $f_{\Phi^\star}(\cdot)$ and to obtain $\mW^\star$. That is, we seek an \ac{nn} with parameters $\Phi$ for which inputs $\tilde{\mY}$ and $\tilde{\mZ}$ result in an output $\mW^\star = f_{\Phi^\star}( \tilde{\mY},\tilde{\mZ} )$ that satisfies~\eqref{prob:form:pow:learn} and~\eqref{prob:form:sinr:learn}.

\subsection{The proposed \ac{nn} architecture}

 \acp{nn} are  designed to work with real-valued data. Therefore, we begin by stacking the real and imaginary parts of $\tilde{\mY}$, one below the other, in the matrix $\tilde{\mY}^{\rm R} \in \mbR^{2M \times K}$, where the $k$th column is given by $\tilde{\vy}^{\rm R}_k =  [ \Re\left( \tilde{\vy}_{k} \right)\rT, \Im\left( \tilde{\vy}_{k} \right)\rT]\rT$. Similarly, let us define $\tilde{\mZ}^{\rm R}$ with the $m$th column $\tilde{\vz}_{m}^{\rm R}  = [ \Re\left( \tilde{\vz}_{m} \right)\rT, \Im\left( \tilde{\vz}_{m} \right)\rT]\rT$. The splitting of real and imaginary parts is carried out by the $\mbC 2 \mbR $ block.  We process $\tilde{\mY}^{\rm R}$ and $\tilde{\mZ}^{\rm R}$ through a series of \acp{mlp} to obtain ${\mW}$. We begin by lifting each vector $\tilde{\vy}_k^{\rm R}$ and $\tilde{\vz}_m^{\rm R}$ to a higher dimension $d > 2M$. To this end, we process $\tilde{\vy}_k^{\rm R}$ with  \texttt{COMM-MLP} $C_{\Phi_{\rm C}}(\cdot) \colon \mbR^{2M} \rightarrow \mbR^{d}$ to obtain $\tilde{\vy}_k^{(1)} = C_{\Phi_{\rm C}}( \tilde{\vy}_{k}^{\rm R} )$ for $k=1,\ldots,K$. Similarly, we process the radar data $\tilde{\vz}_m^{\rm R}$ with \texttt{SENS-MLP} $S_{\Phi_{\rm S}}(\cdot ) \colon \mbR^{2M} \rightarrow \mbR^{d}$ to obtain $\tilde{\vz}_m^{(1)} = S_{\Phi_{\rm S}}( \tilde{\vz}_{m}^{\rm R} )$ for $m=1,\ldots,M$.

Next, we process the higher dimensional representations of the communication~(i.e., $\{ \tilde{\vy}_{k}^{(1)} \}_{k=1}^K$) and sensing data~(i.e., $\{  \tilde{\vz}_{m}^{(1)} \}_{m=1}^M$) using  \texttt{ISAC-MLP}, $I_{\Phi_{\rm I}}(\cdot) \colon \mbR^{d} \rightarrow \mbR^{2M} $ to obtain $K+M$ vectors of length $2M$. Let us collect the output of  \texttt{ISAC-MLP} in a matrix $\tilde{\mW} = I_{\Phi_{\rm I}}( [
	\tilde{\vy}_{1}^{(1)}, \ldots,   \tilde{\vy}_{K}^{(1)}, \tilde{\vz}_{1}^{(1)},  \ldots,  \tilde{\vz}_{M}^{(1)}
	] ) \in \mbR^{2M \times (M+K)}$.
From $\tilde{\mW}$, we construct the $M \times (M+K)$ matrix $\hat{\mW}$ having complex entries using the $\mbR2\mbC$ block as $	\hat{\mW}= [ \tilde{\mW} ]_{[1:M,:]} + \jmath [ \tilde{\mW} ]_{[M+1:2M,:]}$,
where the notation $\left[ \mA \right]_{[n:m,:]}$ refers to the submatrix obtained by collecting the rows of $\mA$ with row indices $n$ to $m$. Finally, we use a normalization layer \texttt{NL} to obtain the transmit precoder $\mW$ that satisfies~\eqref{prob:form:pow:learn} as $\mW = \sqrt{P_{\rm d}}\hat{\mW}/\Vert \hat{\mW}\Vert_F$.
The proposed architecture is summarized in Fig.~\ref{fig:nn:arch}.
Since the dimensions of the associated \acp{mlp} in the proposed \ac{nn} is independent of the system parameters such as $K$, $T$, $L_{\rm p}$, or $L_{\rm r}$, the proposed method is not limited to a given setting and does not require retraining when some or all of these parameters change. Later, through numerical simulations, we demonstrate that the proposed method generalizes well across different test cases.

\vspace{-3mm}
\subsection{The loss function and training}
 {\color{black}To obtain the precoders (i.e., to learn $\Phi^\star$), we propose to train an \ac{nn} in an unsupervised setting, since the optimal precoders~(i.e., the labels) are not known beforehand during training.} Typically,  problems considered in unsupervised settings are unconstrained and the network is trained by minimizing a loss function. For~\eqref{prob:form:learn}, we need to choose a loss function that not only maximizes $Q( f_{\Phi}(  \tilde{\mY}, \tilde{\mZ}) )$ but also satisfy the constraints~\eqref{prob:form:pow:learn} and \eqref{prob:form:sinr:learn}.

 Let us recall that the \texttt{NL} block already ensures that the output satisfies~\eqref{prob:form:pow:learn}.   However, it is not possible to carry out a similar operation on $\mW$ to satisfy the \ac{sinr} constraints~\eqref{prob:form:sinr:learn}. We therefore  develop a loss function that promotes outputs $\mW = f_{\Phi}(\tilde{\mY},\tilde{\mZ})$ that are more likely to satisfy~\eqref{prob:form:sinr:learn}. {\color{black}For convenience, let us define
$ Q(\Phi) \overset{\Delta}{=} Q(f_{\Phi}(\tilde{\mY},\tilde{\mZ}))$ and  $h_k(\Phi) \overset{\Delta}{=} \gamma_k(f_{\Phi}(\tilde{\mY},\tilde{\mZ})) - \Gamma$
 for $k=1,\ldots,K$.}  Then,~\eqref{prob:form:learn} can be rewritten as 
 \begin{equation} \label{prob:mod2} \vspace{-2mm}
 	\underset{\Phi}{\text{maximize}} \,\,  Q\left( \Phi \right)  \quad  \text{s. to} \quad h_k(\Phi) \geq 0, \, k=1,\ldots,K,
 \end{equation}
where we have dropped~\eqref{prob:form:pow:learn} due to \texttt{NL}. To develop the loss function, we begin by computing the first-order optimality conditions of~\eqref{prob:mod2}. The Lagrangian function is given by
\begin{equation} \label{eq:def:lagranginan:main}
	L(\Phi,\boldsymbol{\mu}) = Q(\Phi) + \sum_{k=1}^{K}\mu_k h_k(\Phi),
\end{equation}
where $\boldsymbol{\mu}=[\mu_1,\ldots,\mu_K]\rT$ are the Lagrange multipliers.   At the Karush-Kuhn-Tucker~(KKT) optimal point $(\Phi^\star,\boldsymbol{\mu}^\star)$, we have
\begin{subequations} \label{prob:kkt}
\vspace{-2mm}
	\begin{align} 
		\nabla L(\Phi^\star) &= \boldsymbol{0}, \\
		h_k(\Phi^\star) &\geq 0, \quad \mu_k^\star  \geq 0,  \quad k=1,\ldots,K \label{prob:kkt:c:ineq}  \\
				\sum_{k=1}^{K}\mu_k^\star h_k(\Phi^\star) &= 0 \label{prob:kkt:c:slack}.
	\end{align}
\end{subequations}

 The consequence of~\eqref{prob:kkt:c:ineq} and \eqref{prob:kkt:c:slack} on the Lagrangian can be summarized as follows.  When $h_k(\Phi) > 0$, the $k$th inequality constraint is not active. Hence,~\eqref{prob:kkt:c:slack}  ensures that $\mu_k = 0$, thereby ensuring $\mu_k h_k(\Phi) = 0$. On the other hand, when  $h_k(\Phi) = 0$,  the $k$th inequality constraint is active and  $\mu_k \geq 0$ so that $\mu_k h_k(\Phi) = 0$. Consider the following modification to~\eqref{eq:def:lagranginan:main}, we have 
 \vspace{-2mm}
 \begin{equation}
 	\tilde{L}(\Phi,\boldsymbol{\mu}) = Q(\Phi) +  \sum_{k=1}^{K} \vert \mu_k \vert {\rm max}\left( -h_k(\Phi), 0 \right) h_k(\Phi)^\kappa,
 \end{equation}
where $k$ is an odd number. For a  feasible $(\Phi,\boldsymbol{\mu})$, the behavior of  $\tilde{L}(\Phi,\boldsymbol{\mu})$ is the same as that of the  $L(\Phi,\boldsymbol{\mu})$ since the second term is zero. Whenever the points are not feasible, the second term of $\tilde{L}(\Phi,\boldsymbol{\mu})$  becomes $ -\sum_{k=1}^{K} \vert \mu_k \vert h_k(\Phi)^{\kappa+1} <0$.

To find the first-order optimality point, it is sufficient to learn the optimal values  $\Phi^\star$ and $\boldsymbol{\mu}^\star$. To this end, we propose 
 to train the \ac{nn} with a loss-function  based on $\tilde{L}(\Phi,\boldsymbol{\mu})$, i.e.,
\vspace{-3mm}
\begin{multline}\label{eq:loss:fn:defn:main} 
		\ell_{\Phi,\boldsymbol{\mu}}(\tilde{\mY},\tilde{\mZ}) = \lambda_{\rm S} Q(f_{\Phi}(\tilde{\mY},\tilde{\mZ} )) + \lambda_{\rm C}\sum_{k=1}^{K}\vert \mu_k + 
		 \epsilon \vert \\ \times {\rm max}( -h_k(f_{\Phi}(\tilde{\mY},\tilde{\mZ} )),0 )h_k(f_{\Phi}(\tilde{\mY},\tilde{\mZ} ))^\kappa,
\end{multline}
where $\lambda_{\rm S}$, $\lambda_{\rm C}$, and $\epsilon$ are hyperparameters. We introduced $\lambda_{\rm C}$ and $\lambda_{\rm S}$ to account for the possible scale differences in the two terms in the loss function and  $\epsilon$ is a small number used for numerical stability. In sum, we propose a loss function to find the first-order optimal points $(\Phi^\star,\boldsymbol{\mu}^\star)$ of the constrained optimization problem~\eqref{prob:mod2} by eliminating the constraints and absorbing them in the modified Lagrangian function.

We train the proposed \ac{nn}  model by  minimizing the loss function $-	\ell_{\Phi,\boldsymbol{\mu}}(\tilde{\mY},\tilde{\mZ})$, i.e.,
\begin{equation}
	\Phi^\star,\boldsymbol{\mu}^\star = \underset{\Phi,\boldsymbol{\mu}}{\text{argmin }}  -\mbE\left[ \ell_{\Phi,\boldsymbol{\mu}}(\tilde{\mY},\tilde{\mZ}) \right], \nonumber
\end{equation}
where the expectation is computed over different training examples of $\tilde{\mY}$ and $\tilde{\mZ}$.

\vspace{-3mm}
\subsection{Complexity}
{\color{black}The computational complexity of obtaining $\mW^\star$ using the trained \ac{nn} is as follows. Each layer of an \ac{mlp} consists of a linear transformation followed by an element-wise non-linearity. Let us assume that all the \acp{mlp} comprise of an input layer, an output layer, and $L_{\rm H}$ hidden layers, each of dimension $d_{\rm H}$. Then, processing a $2M$-long vector with $C_{\Phi_{\rm C}}( \cdot )$ or $S_{\Phi_{\rm S}}( \cdot)$ costs about $\cO( 2Md_{\rm H} + L_{\rm H}d_{\rm H}^2 +  d_{\rm H}d )$ flops. Similarly, processing a $d$-long vector using $I_{\Phi_{\rm I}}(\cdot)$ incurs approximately $\cO( dd_{\rm H} + L_{\rm H}d_{\rm H}^2 + 2Md_{\rm H} )$ flops.  The $\mbR2\mbC$ layer followed by \texttt{NL} costs about $\cO(M(M+K))$ flops. Hence, the overall complexity of the \ac{nn}-based solution is approximately $\cO( 2(M+K)( 2Md_{\rm H} + L_{\rm H}d_{\rm H}^2 +  d_{\rm H}d)  )$ flops, which is linear in the number of users $K$. On the other hand, the complexity of the \ac{sdp}-based method is  $\cO( M^{6.5}K^{6.5} \log(1/\delta) )$, which is typically  several orders higher and does not scale well with the number of \acp{ue}.}

	\begin{figure*}[ht!]
	\begin{subfigure}[c]{0.48\columnwidth}\centering
		\includegraphics[width=1\columnwidth]{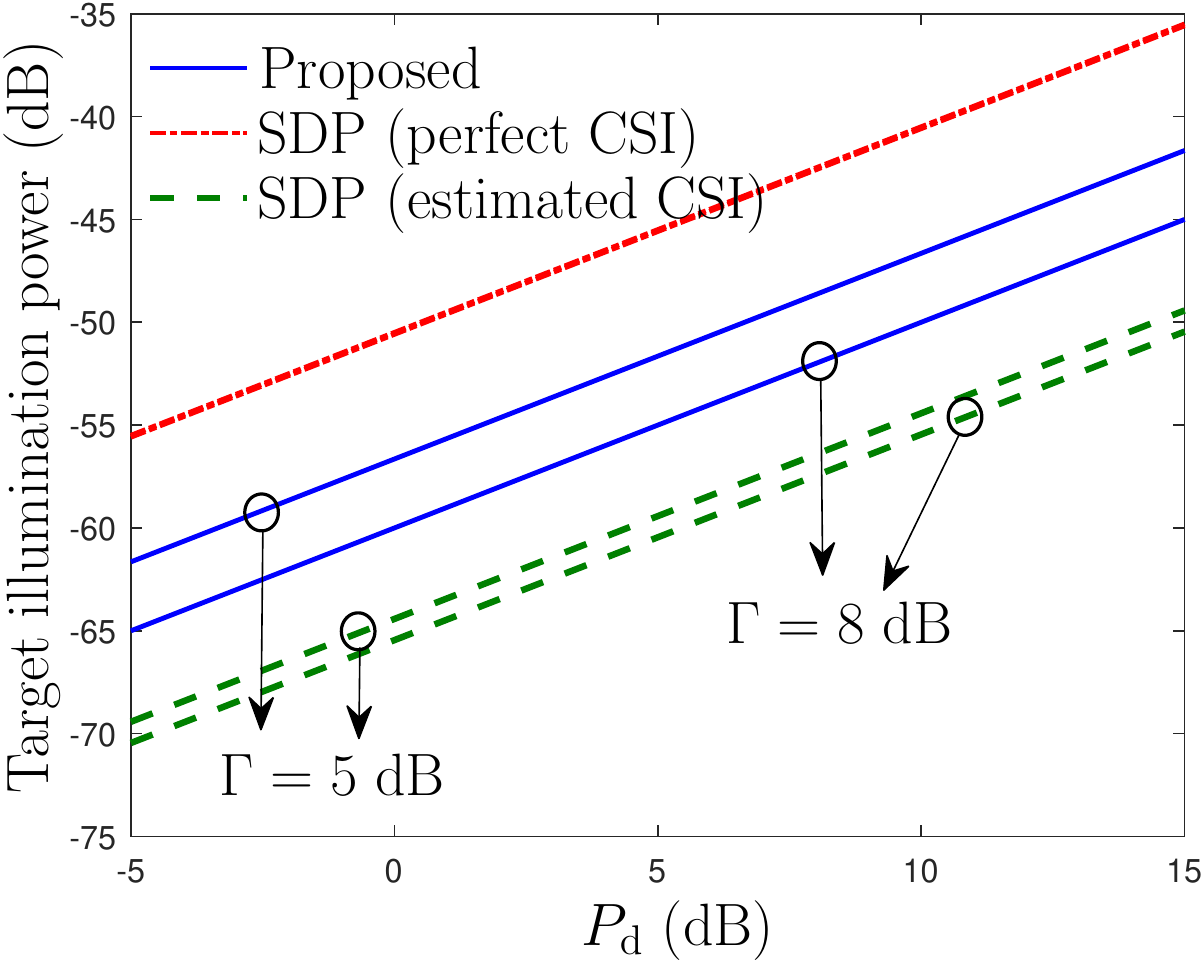}
		\caption{}
		\label{}
	\end{subfigure}
	~
	\begin{subfigure}[c]{0.48\columnwidth}\centering
		\includegraphics[width=1\columnwidth]{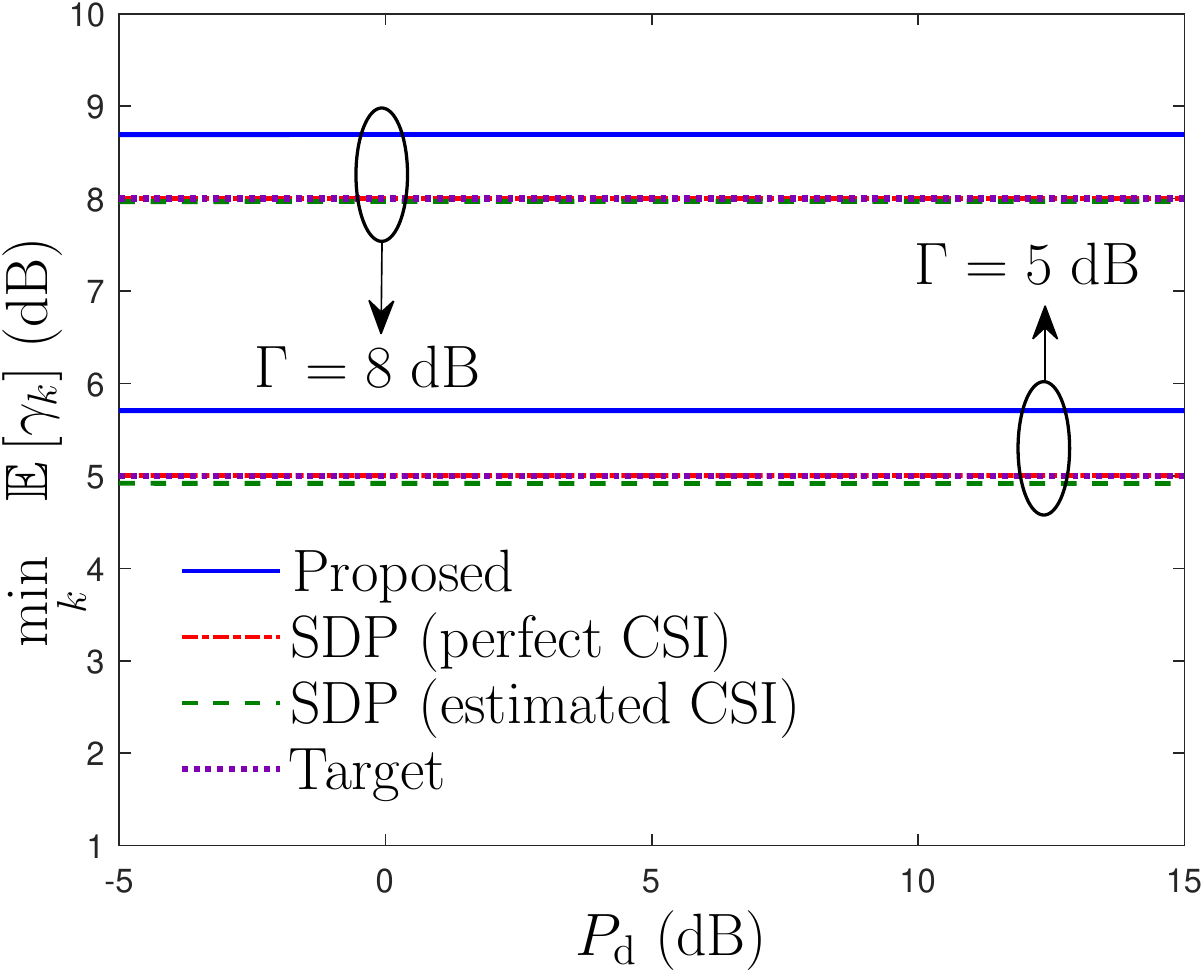}
		\caption{}
		\label{fig:a1_tx}
	\end{subfigure}
	~
	\begin{subfigure}[c]{0.48\columnwidth}\centering
		\includegraphics[width=1\columnwidth]{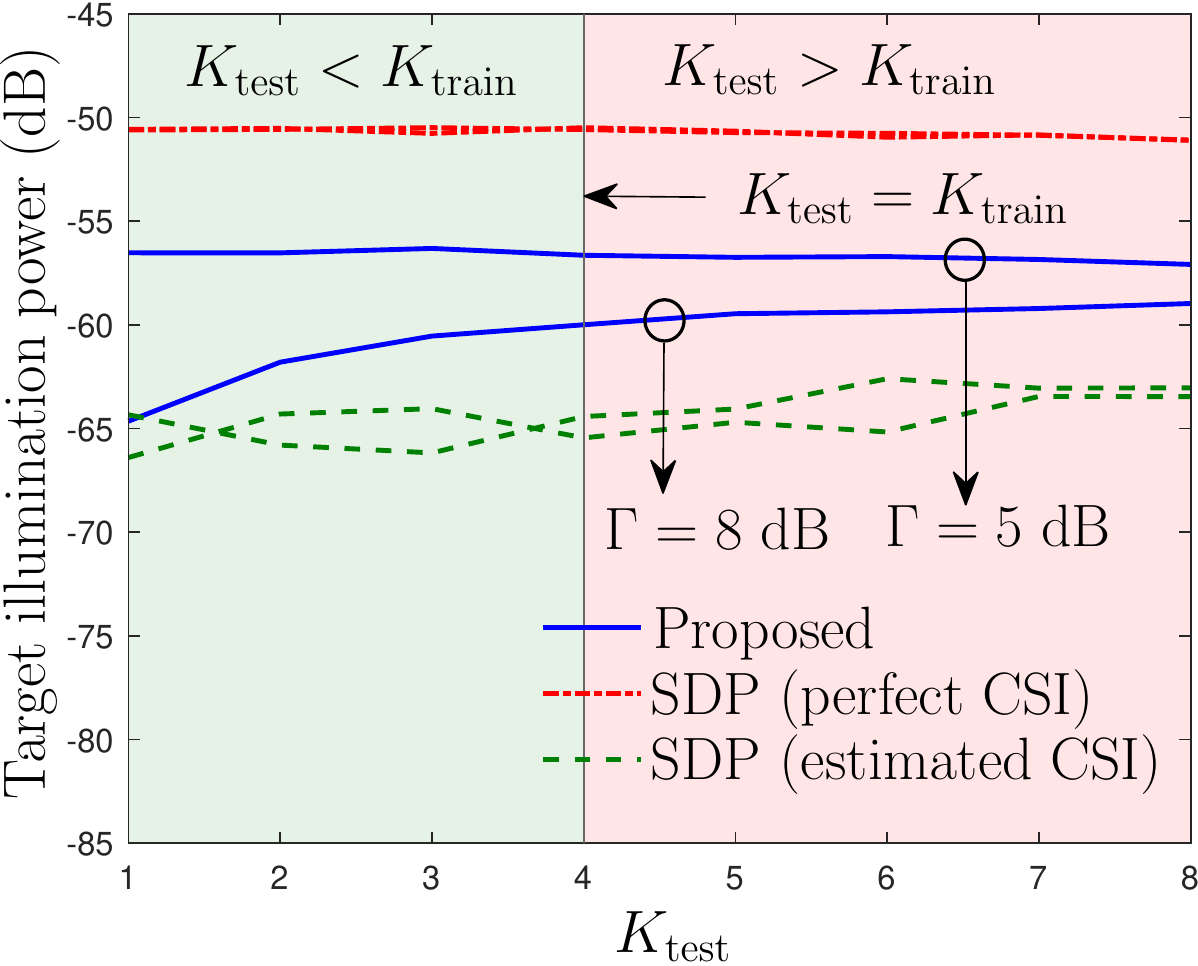}
		\caption{}
		\label{fig:a1_tx_split}
	\end{subfigure}
	~
	\begin{subfigure}[c]{0.48\columnwidth}\centering
		\includegraphics[width=1\columnwidth]{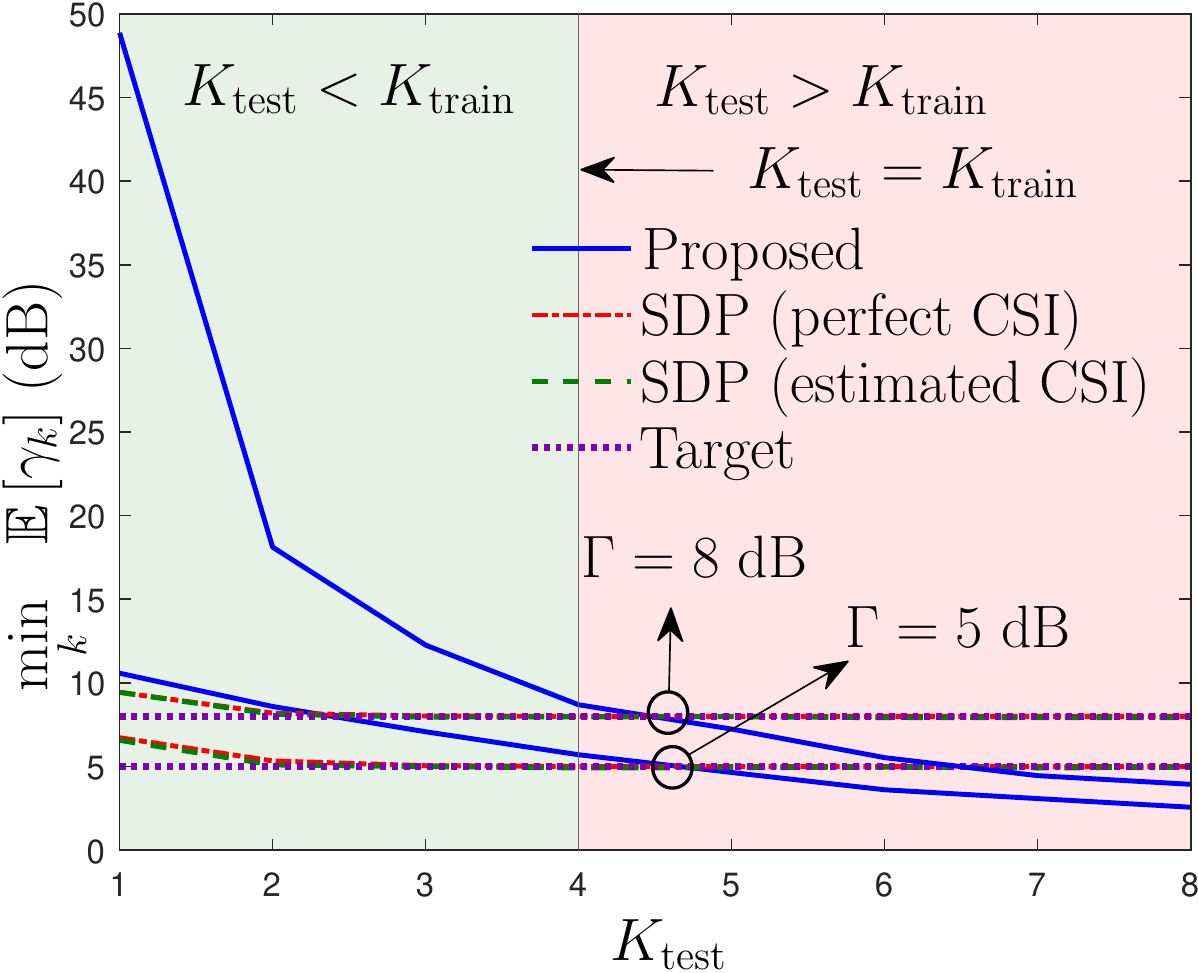}
		\caption{}
		\label{fig:a1_ris}
	\end{subfigure} \vspace{-2mm}
	\caption{ Impact of downlink transmit power $P_{\rm d}$:  (a) Worst-case target illumination power. (b) Worst-case average SINR. Generalizability when the number of users in test set ($K_{\rm test}$) is different from the number of users used in train set ($K_{\rm train}=4$) for $P_{\rm d}=0$ dB: (c) Worst-case target illumination power. (d) Worst-case average SINR.}
	\label{fig:main}
 \vspace{-2mm}
\end{figure*}

\section{Numerical simulations} \label{sec:sim}
We demonstrate the advantages of the proposed method through several numerical simulations. Unless otherwise mentioned, we consider $M=16$, $K=4$, $T=8$, and $\Gamma = 5$ dB. We use communication pilot length and radar echo snapshots as $L_{\rm p} = 20$ and $L_{\rm r} = 32$, respectively. We set transmit powers as $P_{\rm r} = 10$ dB and $P_{\rm u}=0$ dB. The \ac{dfbs} is assumed to be at $(0,0)$m. The users are drawn with co-ordinates $(X,Y)$ where  $X \in [15,18]$m, and $Y \in [8,18]$m. Targets are assumed to be located in a sector between $-80^\circ$ and $-10^\circ$ with a range of $5$ to $20$m. Communication channels are assumed to follow Rayleigh distribution with a pathloss of $30 + 36\log d$ dB, where $d$ is the distance of the user from the \ac{dfbs}. Radar links are assumed to  have a pathloss of $30+22\log d$ dB. The noise variances at the \acp{ue} and at the \ac{dfbs} are selected as $\sigma^2 = -94$ dBm and $\nu^2 = -70$ dBm, respectively. 

{\color{black}Both \texttt{comm-MLP} and \texttt{Sens-MLP}  are two layer \acp{mlp} with the intermediate dimension being $2d$; \texttt{ISAC-MLP}  has $4$ layers with intermediate dimensions $d$, $d$, and $2d$.}
 All layers (except the output layer of \texttt{ISAC-MLP}) use \texttt{ReLU} activation. The output layer of \texttt{ISAC-MLP} is a linear layer. We train the \ac{nn} for $2000$ epochs wherein each epoch comprises of $10$ batches. Each batch consists of $10$   independent realizations of $\tilde{\mY}$ and $\tilde{\mZ}$. We use the hidden dimension as $d=1024$. 
The hyperparameters are selected through grid search as $\lambda_{\rm C}=1$,  $\lambda_{\rm S}=10^7$, $\kappa=3$, and  $\epsilon=10^{-3}$. We implement the proposed \ac{nn} in Pytorch. For training, we use ADAM optimizer with a learning rate of $10^{-4}$. {\color{black}For training, we set $P_{\rm d} = 0$ dB}. We test the \ac{nn} by carrying out inference over $100$ independent channel realizations.
 
We compare the performance of the proposed learning-based solution with that of solving~$(\cP)$ using \ac{sdp}~\cite{liu2020joint_transmit_beamform}. Specifically, we consider two scenarios: in the first, we assume that perfect \ac{csi} is available. We refer to this as \texttt{SDP (perfect CSI)}. In the second scenario, we consider a more realistic situation where the underlying wireless communication channels are estimated using least-squares from $\tilde{\mY}$ and radar channel are estimated using Bartlett beamforming (for estimating $\theta_m$) followed by least squares (for estimating $\{\alpha_m,\beta_m\}$) from $\tilde{\mZ}$. We refer to this as~\texttt{SDP (estimated CSI)}.

 We first evaluate the performance of the proposed method by testing the \ac{nn} on a setting where the statistics of the pilots and echoes are same as the ones used during the training phase. We present the worst-case illumination power of different methods in Fig.~\ref{fig:main}(a). Throughout the considered values of $P_{\rm d}$, the performance of the proposed method is significantly better than that of \texttt{SDP (estimated CSI)}, clearly demonstrating the advantage of using an \ac{nn} to learn the precoder rather than to apply traditional optimization based techniques such as~\cite{liu2020joint_transmit_beamform,sankar2022isac_ris_beamforming} on estimated channels. {\color{black}Even though we used $P_{\rm d} = 0$ dB for training, the proposed \ac{nn} generalizes well across different values of $P_{\rm d}$.} As $\Gamma$ increases to $8$ dB from $5$ dB, $Q$ of \texttt{Proposed} decrease due to more stringent communication constraints. Moreover, due to the presence of noise in $\tilde{\mY}$ and $\tilde{\mZ}$, $Q$ of the \texttt{Proposed} will be inevitably worse than a method using perfect (noiseless) channels. 
 
 Next, we evaluate the communication performance of the proposed method by evaluating the worst-case average \ac{sinr} of the \acp{ue}, $\gamma_{\rm min} = \underset{k}{\text{min}}\quad \mbE\left[ \gamma_k \right]$.  If $\gamma_{\rm min} \geq \Gamma$, we  conclude that the average \ac{sinr} of all users are above $\Gamma$. As we can observe from Fig.~\ref{fig:main}(b), the proposed method ensures that the average \ac{sinr} of all the users are higher than the desired threshold (\texttt{Target}) throughout the considered simulation setting. In other words, we have numerically showed that the proposed \ac{nn}, along with the proposed loss function, succeeds in meeting the communication constraint in a statistical sense~(i.e., on average).

To analyze the generalization capabilities of the proposed \ac{nn}, we now evaluate the trained network on different scenarios, which are different from the ones used for training. We begin by presenting the communication and radar performance when the network is subjected to different \ac{ue} locations in Table~\ref{tab1}.
 \begin{table}[t]  
 \begin{center} 
 	\begin{tabular}{ |c|c|c|c| } 
 		\hline
 		Co-ordinates (m) &  Area (m$^2$) & $\underset{k}{\text{min}} \, \mbE\left[ \gamma_k \right]$ (dB) & $Q$ (dB) \\ 
 		\hline
 		 $X \in [15,18], Y \in [8,18]$  & $30$ & $5.71$ & $-56.73$  \\
 		 $X \in [10,20], Y \in [5,25]$  & $200$ & $6.33$ & $-56.65$  \\
 		  $X \in [5,25], Y \in [5,25]$  & $400$ & $6.81$ & $-56.91$  \\
 		\hline
 	\end{tabular}
 \end{center}
\caption{Generalizability of proposed model for different user locations $(X,Y)$ and $\Gamma = 5$ dB. }
\vspace{-5mm}
\label{tab1}
\end{table}
The network generalizes well across different \ac{ue} locations and provide consistent results even when the test area is around $10$ times larger than the locations for which the network is trained for. Next, we evaluate the performance of the proposed method on a scenario where the number of users are different in the training and testing phases. Specifically, let $K_{\rm train}=4$ and $K_{\rm test}$ denote the number of \acp{ue} during the training and testing phases, respectively. In Fig.~\ref{fig:main}(c) and Fig.~\ref{fig:main}(d), we present the worst-case target illumination power and the worst-case average \ac{sinr} of the users, respectively, for different values of $K_{\rm test}$. As before, the proposed scheme clearly outperforms \texttt{SDP (estimated CSI)} throughout the considered range in terms of the sensing performance.  Interestingly, the communication performance of \texttt{Proposed} is infact better whenever $K_{\rm test} < K_{\rm train}$ since the system is subjected to a simpler setting (simpler since the multi-user interference decreases when the number of users decrease) during testing than the one used during training. In general, the worst-case average \ac{sinr} of \texttt{Proposed} decreases with an increase in $K_{\rm test}$. While $\gamma_{\rm min}< \Gamma$ for $K_{\rm test} > K_{\rm train}$, it is important to recall that we have to re-run the benchmark schemes from scratch whenever there is a change in the number of users and that the complexity of the \ac{sdp}-based solution grows as $K^{6.5}$. On the other hand, the complexity of the proposed method scales linearly with the number of users (i.e, as $K$ vs $K^{6.5}$), making it much suited for next-generation massive MIMO systems.

\section{Conclusions} \label{sec:conclusion}
In this paper,  we proposed an \ac{nn} approach to learn the transmit precoders from the received echo signals and pilots at the \ac{dfbs} while avoiding the need for explicit channel estimation. The transmit precoders are designed to maximize the worst-case target illumination power while guaranteeing a prescribed SINR for the users on average. We develop a loss function based on first-order optimality conditions to train the NN model in an unsupervised setting. Through numerical simulations, we demonstrate that the proposed method outperforms traditional SDP-based methods in presence of channel estimation errors and it incurs lower computational complexity.

	\bibliographystyle{IEEEtran}
	\bibliography{IEEEabrv,bibliography}

\end{document}

%% file: macros.tex
\def\cast{{
   \mathord{
      \hbox to 0em{
         \ooalign{
	   \smash{\hbox{$\ast$}}\crcr
	   \smash{\hskip-1pt\Large\hbox{$\circ$}} }
	 \hidewidth}
      \phantom{\bigcirc}
} }}

\newcommand{\rH}{^{ \raisebox{1pt}{$\rm \scriptscriptstyle H$}}}
\newcommand{\rT}{^{ \raisebox{1.2pt}{$\rm \scriptstyle T$}}}

\newcommand{\bds}{\begin {itemize}}
\newcommand{\eds}{\end {itemize}}
\newcommand{\bdf}{\begin{definition}}
\newcommand{\blm}{\begin{lemma}}
\newcommand{\edf}{\end{definition}}
\newcommand{\elm}{\end{lemma}}
\newcommand{\bthm}{\begin{theorem}}
\newcommand{\ethm}{\end{theorem}}
\newcommand{\bprp}{\begin{prop}}
\newcommand{\eprp}{\end{prop}}
\newcommand{\bcl}{\begin{claim}}
\newcommand{\ecl}{\end{claim}}
\newcommand{\bcr}{\begin{coro}}
\newcommand{\ecr}{\end{coro}}
\newcommand{\bquest}{\begin{question}}
\newcommand{\equest}{\end{question}}

\newcommand{\larrow}{{\larrow}}

\newcommand{\argmin}{\ensuremath{\mathrm{arg}\min}}
\newcommand{\argmax}{\ensuremath{\mathrm{arg}\max}}

\newcommand{\cC}{{\ensuremath{\mathcal{C}}}}

\newcommand{\cN}{{\ensuremath{\mathcal{N}}}}
\newcommand{\cO}{{\ensuremath{\mathcal{O}}}}
\newcommand{\cP}{{\ensuremath{\mathcal{P}}}}

\def\mbC{{\ensuremath{\mathbb C}}}

\def\mbE{{\ensuremath{\mathbb E}}}

\def\mbR{{\ensuremath{\mathbb R}}}

\newcommand{\va}{{\ensuremath{{\mathbf{a}}}}}

\newcommand{\vc}{{\ensuremath{{\mathbf{c}}}}}
\newcommand{\vd}{{\ensuremath{{\mathbf{d}}}}}

\newcommand{\vg}{{\ensuremath{{\mathbf{g}}}}}

\newcommand{\vh}{{\ensuremath{{\mathbf{h}}}}}

\newcommand{\vs}{{\ensuremath{{\mathbf{s}}}}}

\newcommand{\vt}{{\ensuremath{{\mathbf{t}}}}}

\newcommand{\vv}{{\ensuremath{{\mathbf{v}}}}}

\newcommand{\vx}{{\ensuremath{{\mathbf{x}}}}}

\newcommand{\vy}{{\ensuremath{{\mathbf{y}}}}}
\newcommand{\vz}{{\ensuremath{{\mathbf{z}}}}}

\newcommand{\mA}{{\ensuremath{\mathbf{A}}}}

\newcommand{\mC}{{\ensuremath{\mathbf{C}}}}

\newcommand{\mE}{{\ensuremath{\mathbf{E}}}}
\newcommand{\mF}{{\ensuremath{\mathbf{F}}}}

\newcommand{\mH}{{\ensuremath{\mathbf{H}}}}

\newcommand{\mI}{{\ensuremath{\mathbf{I}}}}

\newcommand{\mN}{{\ensuremath{\mathbf{N}}}}

\newcommand{\mS}{{\ensuremath{\mathbf{S}}}}

\newcommand{\mV}{{\ensuremath{\mathbf{V}}}}

\newcommand{\mW}{{\ensuremath{\mathbf{W}}}}

\newcommand{\mY}{{\ensuremath{\mathbf{Y}}}}
\newcommand{\mZ}{{\ensuremath{\mathbf{Z}}}}

\usepackage{latexsym}

\def\IC{\mathbb C}
\def\IN{\mathbb N}
\def\IZ{\mathbb Z}
\def\IR{\mathbb R}

\def\shat{^{\mathchoice{}{}%
 {\,\,\smash{\hbox{\lower4pt\hbox{$\widehat{\null}$}}}}%
 {\,\smash{\hbox{\lower3pt\hbox{$\hat{\null}$}}}}}}

\def\bSigma{{
      \ooalign{
      \smash{\hskip.4pt\raise.4pt\hbox{$\Sigma$}}\vphantom{}\crcr
      \smash{\hskip.7pt\raise.6pt\hbox{$\Sigma$}}\vphantom{}\crcr
      \smash{\hbox{$\Sigma$}}\vphantom{$\Sigma$}}
      \vphantom{\hbox{$\Sigma$}}
      }}
\def\bTheta{{
      \ooalign{
      \smash{\hskip.5pt\raise.5pt\hbox{$\Theta$}}\vphantom{}\crcr
      \smash{\hskip.0pt\raise.1pt\hbox{$\Theta$}}\vphantom{}\crcr
      \smash{\hbox{$\Theta$}}\vphantom{$\Theta$}}
      \vphantom{\hbox{$\Theta$}}
      }}
\def\bDelta{{
      \ooalign{
      \smash{\hskip.4pt\raise.4pt\hbox{$\Delta$}}\vphantom{}\crcr
      \smash{\hskip.7pt\raise.6pt\hbox{$\Delta$}}\vphantom{}\crcr
      \smash{\hbox{$\Delta$}}\vphantom{$\Delta$}}
      \vphantom{\hbox{$\Delta$}}
      }}
\def\bLambda{{
      \ooalign{
      \smash{\hskip.5pt\raise.5pt\hbox{$\Lambda$}}\vphantom{}\crcr
      \smash{\hskip.0pt\raise.1pt\hbox{$\Lambda$}}\vphantom{}\crcr
      \smash{\hbox{$\Lambda$}}\vphantom{$\Lambda$}}
      \vphantom{\hbox{$\Lambda$}}
      }}

\makeatletter

\def\bordermatrix#1{\begingroup \m@th
  \@tempdima 8.75\p@
  \setbox\z@\vbox{%
    \def\cr{\crcr\noalign{\kern2\p@\global\let\cr\endline}}%
    \ialign{$##$\hfil\kern2\p@\kern\@tempdima&\thinspace\hfil$##$\hfil
      &&\quad\hfil$##$\hfil\crcr
      \omit\strut\hfil\crcr\noalign{\kern-\baselineskip}%
      #1\crcr\omit\strut\cr}}%
  \setbox\tw@\vbox{\unvcopy\z@\global\setbox\@ne\lastbox}%
  \setbox\tw@\hbox{\unhbox\@ne\unskip\global\setbox\@ne\lastbox}%
  \setbox\tw@\hbox{$\kern\wd\@ne\kern-\@tempdima\left[\kern-\wd\@ne
    \global\setbox\@ne\vbox{\box\@ne\kern2\p@}%
    \vcenter{\kern-\ht\@ne\unvbox\z@\kern-\baselineskip}\,\right]$}%
  \null\;\vbox{\kern\ht\@ne\box\tw@}\endgroup}
\makeatother

\makeatletter
\def\argmin{\mathop{\operator@font arg\,min}}
\def\argmax{\mathop{\operator@font arg\,max}}
\makeatother

\newcommand{\bea}{\begin{array}}
\newcommand{\ena}{\end{array}}
\newcommand{\beq}{\begin{equation}}
\newcommand{\enq}{\end{equation}}

\newcommand{\beqa}{\begin{eqnarray}}
\newcommand{\enqa}{\end{eqnarray}}

\newcommand{\beqan}{\begin{eqnarray*}}
\newcommand{\enqan}{\end{eqnarray*}}

\newcommand{\AL}{\begin{enumerate}}
\newcommand{\ALE}{\end{enumerate}}

\def\addots{\mathinner{
    \mkern1mu\raise0pt\vbox{\kern7pt\hbox{.}}
    \mkern2mu\raise4pt\hbox{.}
    \mkern2mu\raise7pt\hbox{.}
    \mkern1mu}}

\def\sddots{\mathinner{
    \mkern.8mu\raise7pt\hbox{.}
    \mkern.8mu\raise4pt\hbox{.}
    \mkern.8mu\raise0pt\vbox{\kern7pt\hbox{.}}
    \mkern1mu}}

\def\saddots{\mathinner{
    \mkern.2mu\raise0pt\vbox{\kern7pt\hbox{.}}
    \mkern.2mu\raise4pt\hbox{.}
    \mkern.2mu\raise7pt\hbox{.}
    \mkern1mu}}

\def\sqplus{\mathbin{
	{\ooalign{\hfil\raise.3ex\hbox{\scriptsize
	+}\hfil\crcr\mathhexbox274\crcr\mathhexbox275}}
	}} 
\def\sqminus{\mathbin{
	{\ooalign{\hfil\raise.3ex\hbox{\scriptsize
	--}\hfil\crcr\mathhexbox274\crcr\mathhexbox275}}
	}}

\def\IC{{
   \mathord{
      \hbox to 0em{
	 \hskip-4pt
         \ooalign{
	   \smash{\hskip1.9pt\raise2.6pt\hbox{$\scriptscriptstyle |$}}\crcr
	   \smash{\hbox{\rm\sf C}} }
	 \hidewidth}
      \phantom{\hbox{\rm\sf C}}
} }}
\def\IN{
    {\ooalign{
   \smash{\hskip2.2pt\raise1.5pt\hbox{$\scriptscriptstyle |$}}\vphantom{}\crcr
   \hbox{\sf N}
	}}
	} 
\def\IZ{
    {\ooalign{
   \smash{\hskip1.9pt\raise0pt\hbox{$\sf Z$}}\vphantom{}\crcr
   \hbox{\sf Z}
	}}
	} 
\def\IR{
    {\ooalign{
   \smash{\hskip2.2pt\raise1.5pt\hbox{$\scriptscriptstyle |$}}\vphantom{}\crcr
   \smash{\hskip2.2pt\raise3.3pt\hbox{$\scriptscriptstyle |$}}\vphantom{}\crcr
   \hbox{\sf R}
	}}
	} 

\DeclareMathAlphabet{\mathcmb}{OT1}{cmr}{b}{n}

\def\bSigma{\ensuremath{\mathcmb{\Sigma}}}
\def\bLambda{\ensuremath{\mathcmb{\Lambda}}}

\def\bTheta{\ensuremath{\mathcmb{\Theta}}}

\newcommand{\SI}{\begin{indlist}}
\newcommand{\EI}{\end{indlist}}

\newcommand{\DL}{\begin{dashlist}}
\newcommand{\DLE}{\end{dashlist}}

\makeatletter
\def\setboxz@h{\setbox\z@\hbox}
\def\wdz@{\wd\z@}
\def\boxz@{\box\z@}
\def\underset#1#2{\binrel@{#2}%
  \binrel@@{\mathop{\kern\z@#2}\limits_{#1}}}
\def\binrel@#1{\begingroup
  \setboxz@h{\thinmuskip0mu
    \medmuskip\m@ne mu\thickmuskip\@ne mu
    \setbox\tw@\hbox{$#1\m@th$}\kern-\wd\tw@
    ${}#1{}\m@th$}%
  \edef\@tempa{\endgroup\let\noexpand\binrel@@
    \ifdim\wdz@<\z@ \mathbin
    \else\ifdim\wdz@>\z@ \mathrel
    \else \relax\fi\fi}%
  \@tempa
}
\let\binrel@@\relax%
\makeatother
